\documentclass[aps,prl,showpacs,twocolumn,superscriptaddress,floatfix]{revtex4-1}
\usepackage{graphicx,graphics}
\usepackage{dcolumn}
\usepackage{amsmath,amssymb,amsfonts}
\usepackage{latexsym,verbatim}
\usepackage{bm}
\usepackage{color}
\usepackage{ulem}
\usepackage[percent]{overpic}
\usepackage[breaklinks=true,colorlinks,citecolor=blue,linkcolor=blue,urlcolor=blue]{hyperref}
\usepackage{tikz}
\usetikzlibrary{decorations.markings}
\usepackage{mathtools}
\usepackage{subfig}

\newcommand{\e}{\text{e}}
\begin{document}
\title{Tuning the Drude weight of  Dirac-Weyl fermions in one-dimensional ring traps}
\author{Manon Bischoff}
\thanks{These authors contributed equally to this work.}
\affiliation{Johannes Gutenberg-Universit\"{a}t, Institut f\"{u}r Physik, Staudingerweg~7, 55128 Mainz, Germany}
\email{manon.bischoff@uni-mainz.de}
\author{Johannes J\"unemann}
\thanks{These authors contributed equally to this work.}
\affiliation{Johannes Gutenberg-Universit\"{a}t, Institut f\"{u}r Physik, Staudingerweg~7, 55128 Mainz, Germany}
\affiliation{Graduate School Materials Science in Mainz, Staudingerweg~9, 55128 Mainz, Germany}
\email{johannes.juenemann@uni-mainz.de}
\author{Marco Polini}
\affiliation{Istituto Italiano di Tecnologia, Graphene Labs, Via Morego~30, I-16163 Genova, Italy}
%
\author{Matteo Rizzi}
\affiliation{Johannes Gutenberg-Universit\"{a}t, Institut f\"{u}r Physik, Staudingerweg~7, 55128 Mainz, Germany}

%
\begin{abstract}
We study the response to an applied flux of an interacting system of Dirac-Weyl fermions confined in a one-dimensional (1D) ring. Combining analytical calculations with density-matrix renormalization group results, we show that tuning of interactions leads to a unique many-body system that displays either a suppression or an enhancement of the Drude weight---the zero-frequency peak in the ac conductivity---with respect to the non-interacting value.  An asymmetry in the interaction strength between same- and different-pseudospin Dirac-Weyl fermions leads to Drude weight enhancement. Viceversa, symmetric interactions lead to Drude weight suppression. Our predictions can be tested in mixtures of ultracold fermions in 1D ring traps.
\end{abstract}

\maketitle 

\noindent {\it Introduction.---}Quantum many-particle systems confined in one-dimensional (1D) ring geometries have attracted a great deal of attention since B\"{u}ttiker, Imry, and Landauer~\cite{buttiker_physlett_1983} showed that they can support persistent currents (PCs) akin to those flowing in superconductors~\cite{Viefers04}. Although they lead to tiny signals, which are extremely susceptible to the environment, PCs have been observed in resistive metallic rings~\cite{bleszynski_science_2009}. 

Ultracold atoms confined in ring-shaped traps~\cite{Sauer01,Gupta05,Ryu07,Lesanovsky07} offer an entirely new class of quantum many-body systems where the orbital response to an applied flux can be studied as a function of temperature, statistics, and interactions, in the absence of extrinsic effects. In these setups, PCs can be launched by rotating a localized barrier~\cite{Ramanathan11,Wright13} or imprinting a suitable artificial gauge field by optical means~\cite{Cooper10,Dalibard11,Goldman14}. Intriguing phenomena 
such as quantization~\cite{Eckel14a}, hysteresis~\cite{Eckel14b}, and decay~\cite{Beattie13} of PCs have already been observed in single or multi-component gases. 
Applications to high-precision measurements, atom interferometry, and quantum information 
can be envisioned
and have spurred the field of ``atomtronics''~\cite{Ryu13,Aghamalyan15}. 
Noticeably, recent advances have made it possible to move towards a strictly 1D regime~\cite{SinucoLeon14}, with optical lattices that can, at least in principle, be added along the ring~\cite{Amico05,Lacki16}, for achieving strong correlations.

These studies have focused on the orbital response of ordinary Schr\"{o}dinger (massive, non-relativistic) fermions and bosons. It is however well known that systems behaving as ultra-relativistic Dirac-Weyl fermions harbor an intriguing orbital response~\cite{MCClure56,MCClure60,MCClure74,Koshino10,principi_prl_2010,gomez_prl_2011,raoux_prl_2014,piechon_prb_2016,gutierrez_prb_2016}, which is sensitive to many-body effects~\cite{principi_prl_2010} even in the absence of impurities because of the intrinsic lack of Galilean invariance~\cite{abedinpour_prb_2011}. In the case of two-dimensional (2D) Dirac-Weyl fermions, which can be found e.g. in graphene~\cite{geim_naturemater_2007} and 2D artificial honeycomb lattices~\cite{polini_naturenano_2013}, McClure predicted long-time ago a very large diamagnetic response at half filling~\cite{MCClure56,MCClure60,MCClure74,Koshino10}. Away from half filling the situation is delicate and ``high-energy'' lattice effects~\cite{gomez_prl_2011} together with many-body effects~\cite{principi_prl_2010} provide interesting scenarios and, possibly, a many-body paramagnetic response~\cite{principi_prl_2010}.

\begin{figure}[t]
\centering
\includegraphics[width=0.8\linewidth]{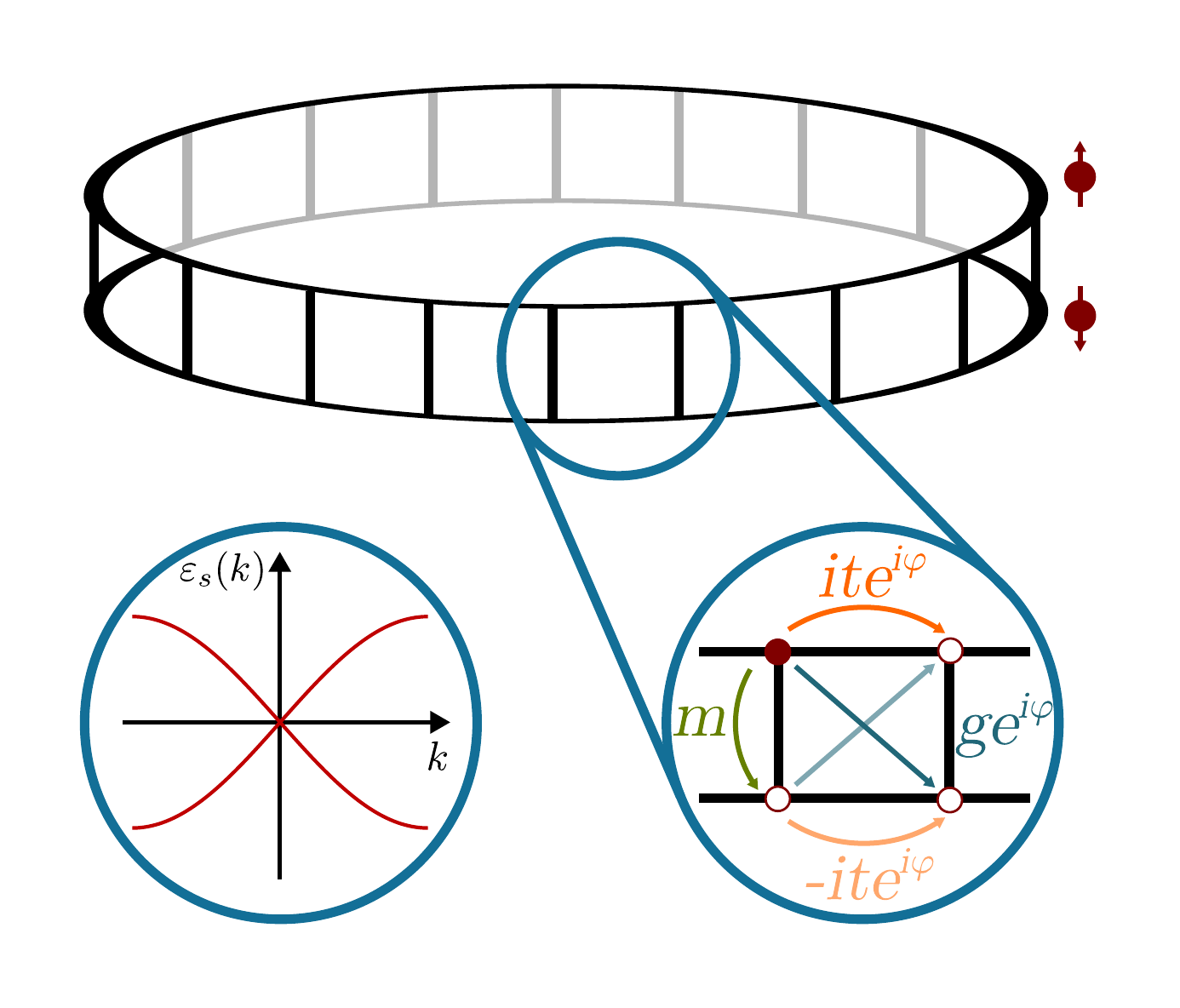}
\caption{(Color online) A sketch of the Creutz ladder model (\ref{eq:CL}) for fermions endowed with a pseudospin degree of freedom $\sigma=\uparrow,\downarrow$. The right inset illustrates all the single-particle terms in the Hamiltonian (\ref{eq:CL}), while the left inset illustrates the Dirac-Weyl crossing in the energy-quasimomentum dispersion $\varepsilon_{s}(k)$ for $t=m=g$ and both $s=\pm$ bands.\label{CreutzLadder}}
\end{figure}

Leaving aside graphene~\cite{geim_naturemater_2007}, its artificial analogues~\cite{polini_naturenano_2013}, and recently discovered Weyl semimetals~\cite{Hasan_2017,Yan_2017}, Dirac-Weyl fermions are also available in cold atom setups~\cite{tarruell_nature_2012,jotzu_nature_2014} and 1D edge states of two-dimensional topological insulators~\cite{hasan_rmp_2010,qi_rmp_2011}. Motivated by this body of literature, we here present a theoretical study of the orbital response of 1D Dirac-Weyl fermions. Specifically, we present extensive calculations of the Drude weight $D$ of 1D Dirac-Weyl fermions in a ring geometry. This quantity controls~\cite{Kohn64,Fye91,Scalapino92} the behavior of the real and imaginary parts of the ac conductivity $\sigma(\omega)$ in the low-frequency $\omega \to 0$ regime. Kohn~\cite{Kohn64} was the first to understand that, for 1D systems with periodic boundary conditions (PBCs), $D$ can be calculated from the dependence of the ground-state energy $E$ on the flux $\Phi$ threading the ring. Our approach treats on equal footing low-energy Dirac-Weyl physics, high-energy lattice effects, and inter-particle interactions. The latter, in particular, are handled in a non-perturbative fashion via a tensor-network generalization of the density-matrix renormalization group (DMRG) approach~\cite{DMRG,cominotti_prl_2014}. We employ binary-tree tensor networks, since they can handle PBCs as efficiently as open boundary conditions~\cite{Gerster14,Gerster16}. These numerical calculations are complemented by diagrammatic many-body perturbation theory. We find that tuning the asymmetry between same- and different-(pseudo)spin Dirac-Weyl fermion interactions can result in qualitative changes of the Drude weight with respect to the non-interacting value $D_{0}$. Asymmetric (symmetric) interactions lead to Drude weight enhancement (suppression). 
We argue that ultracold 1D Dirac-Weyl fermions in ring traps allow for the observation of our predictions.
In passing, we note that the authors of Ref.~\onlinecite{sticlet_prb_2013} calculated $D_{0}$ for 1D Dirac-Weyl fermions, while first-order interaction corrections were calculated~\cite{abedinpour_prb_2011} for the case of graphene.

{\it Interacting Creutz ladder models.---}In order to model 1D fermions with a Dirac-Weyl crossing, we use the Creutz ladder Hamiltonian~\cite{Creutz99}:
\begin{equation}\label{eq:CL}
{\cal H}_0= \frac{1}{2} \sum_{j=1}^L \left[ c^\dagger _j ( it\sigma_3-g\sigma_1) c_{j-1}+ c^\dagger_j m\sigma_1 c_j\right]+ \text{H.c.}~.
\end{equation}
Here, $L$ denotes the number of lattice sites, which we assume to be even throughout this work, $\sigma_{1}$ and $\sigma_{3}$ are ordinary $2 \times 2$ Pauli matrices, and the fermionic operators are two-component spinors $c_j=(c_{j,\uparrow},c_{j,\downarrow})^{\rm T}$, where $\sigma=\uparrow,\downarrow$ refers to any two-valued pseudospin degree of freedom. The model consists of two chains, each representing one of the two pseudospins, 
with different nearest-neighbor (intra-chain $t$ and inter-chain $g$)  hopping parameters and a local spin-flip term ($m$)---see Fig.~\ref{CreutzLadder}. 
For $t=m=g$, the energy-quasimomentum dispersion shows a single Dirac-Weyl crossing ($\hbar=1$ throughout this manuscript) $\varepsilon(k) = \pm v_{\rm F} k$ located at $k=0$---see Fig.~\ref{CreutzLadder}---i.e.~it does not suffer from the fermion doubling problem.

We consider PBCs, which can be realized by closing the 1D system in a ring geometry.
We introduce the Fourier transform of the fermionic operators,
 $c_k =1/\sqrt{L} \sum_{j=1}^{L}e^{-i ka j} c_{j}$, where $a$ is the lattice constant. 
 The values of $k$ are bound to lie in the first Brillouin zone (BZ), $k = 2\pi m/(La) $ where $m \in [-L/2, L/2-1]$ is an integer. A magnetic flux $\Phi$ threading the ring results in a twist of the boundary conditions
or---gauge-equivalently---in a uniform vector potential along the ring.
This appears as a Peierls phase $\varphi = (2\pi/L) (\Phi/\Phi_{0})$ with $0 \leq \Phi \leq \Phi_0$ picked up by the fermions hopping from one site to a neighboring one. Here, $\Phi_{0}$ is the flux quantum, and we set henceforth $a=\Phi_{0}=1$.
The phase $\varphi$ results in a shift $\tilde{k}=k+\varphi$ of the momenta of the Hamiltonian:
\begin{equation}\label{eq:CL_FT_gauge}
{\cal H}_0= t \sum_{k}  c^\dagger_k \left\{\sigma_3\sin (a\tilde{k}) + [1-\cos (a\tilde{k})]\sigma_1\right\} c_k.
\end{equation}
In order to diagonalize ${\cal H}_{0}$, we introduce the operators $d^\dagger_{k,s}=\sum_{\sigma} N^\sigma_{\tilde{k},s} c^\dagger_{k,\sigma}$ and $d_{k,s}=\sum_{\sigma} N^\sigma_{\tilde{k},s} c_{k,\sigma}$, where $s=\pm$ is a band index ($s=-$ for the valence band, $s=+$ for the conduction band) and the form factors $N^{\uparrow}_{k,s} = s g_{k}/\sqrt{2(1- sf_{k})}$ and $N^{\downarrow}_{k,s} = (1- sf_{k})/\sqrt{2(1- sf_{k})}$, with $g_{k} =\text{sgn}(k) \sin(k/2)$ and $f_{k} =\text{sgn}(k)\cos(k/2)$. In this basis we find ${\cal H}_0=-2t \sum_{k,s} s g_{\tilde{k}}~d^\dagger_{k,s} d_{k,s}$. The resulting energy dispersion $\varepsilon_{s}(k)=-2ts g_{\tilde{k}}$ for $s=\pm$, $\Phi=0$, and $t=m=g$ is shown in Fig.~\ref{CreutzLadder}.

We now add inter-particle interactions to the free model (\ref{eq:CL}). These are usually written in the form ${\cal H}_{1}=\sum_{\sigma,\sigma'}\sum_{i,j} v^{\sigma,\sigma'}_{ij} n_{i,\sigma}n_{j,\sigma'}$, where 
$n_{i, \sigma} = c^\dagger_{i,\sigma}c_{i,\sigma}$ is the density operator at site $i$, for pseudospin flavor $\sigma$. In this work we consider the following interactions:
\begin{equation}\label{eq:interactions}
v^{\sigma,\sigma'}_{ij} = \frac{1}{2}(U\delta_{i,j}\delta_{\bar{\sigma},\sigma'} + V\delta_{j,i\pm 1}\delta_{\sigma,\sigma'} + \tilde{V} \delta_{j,i\pm 1}\delta_{\bar{\sigma},\sigma'} )~. 
\end{equation}
Here, $\bar{\sigma}$ is the opposite of $\sigma$. The first term in round brackets on the right-hand side of Eq.~(\ref{eq:interactions}) is an on-site Hubbard-$U$ interaction term acting only between different-pseudospin fermions. The second and third terms are nearest-neighbor interactions acting between same- and different-pseudospin fermions, respectively. The Fourier transform of the interaction $v^{\sigma,\sigma'}_{ij}=(1/L)\sum_k v^{\sigma,\sigma'}_{q} \e^{-iq(i-j)}$
is $v^{\sigma,\sigma'}_{q} = [U \delta_{\bar{\sigma},\sigma'}/2+V\cos(q)\delta_{\sigma,\sigma'} + \tilde{V}\cos(q)\delta_{\bar{\sigma},\sigma'}]$.
\begin{figure}[t]
\includegraphics[width=0.80\linewidth]{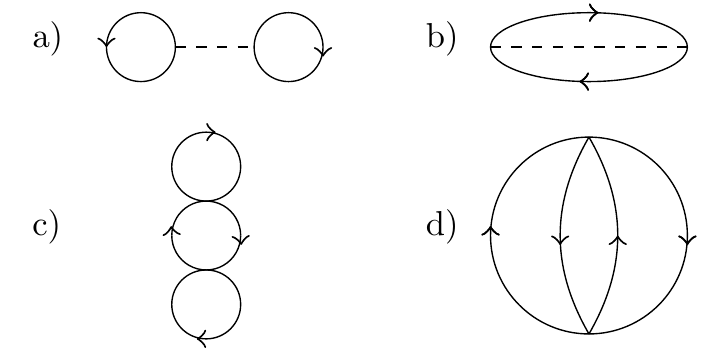}
\caption{First- and second-order perturbation theory diagrams. 
a) Hartree diagram; b) Fock diagram; c) and d) second-order Hugenholtz diagrams~\cite{NegeleOrland}. Solid lines denote bare propagators in all diagrams. Dashed lines in panels a) and b) denote inter-particle interactions.\label{HartreeFock}}
\end{figure}
In the band representation, we find
\begin{eqnarray}\label{Interaction}
{\cal H}_{1} &=& \frac{1}{L}\sum_{s,s'; r,r'}\sum_{k,k',q}  W^{s r r' s'}_{k k' q} d^\dagger_{k+q,s}d^\dagger_{k'-q,r}d_{k',r'}d_{k,s'}~,
\end{eqnarray}
where $s,s',r$, and $r'$ are band indices and 
$W^{s r r' s'}_{k k' q}  = \sum_{\sigma,\sigma'} v^{\sigma,\sigma'}_{q} N^\sigma_{k+q,s} N^{\sigma'}_{k'-q,r} N^{\sigma'}_{k',r'} N^\sigma_{k,s'}$.
Results for the Drude weight in the presence of interactions {\it qualitatively} depend on: 1) finiteness of nearest-neighbor interactions and 2) asymmetry between same- and different-pseudospin nearest-neighbor interactions, which is quantified by the difference $V-\tilde{V}$.

\noindent {\it Drude weight, many-body perturbation theory, and non-perturbative numerical analysis.---}The Drude weight $D$ is formally defined by ${\rm Re}[\sigma(\omega)] = D \delta(\omega) + \sigma_{\rm s}(\omega)$, where $\sigma(\omega)$ is the ac conductivity~\cite{Pines_and_Nozieres} and $\sigma_{\rm s}(\omega)$ is a smooth contribution at $\omega \neq 0$. In 1D systems, $D$ is given by the second derivative of the ground-state energy $E(\Phi)$ with respect to the applied flux~\cite{Kohn64,paritynote},
\begin{equation}
D=\frac{L}{2\pi} \left. \frac{\partial^2 E}{\partial \Phi^2} \right|_{\Phi=0}~.
\end{equation}
For the Creutz ladder model (\ref{eq:CL}) in the absence of inter-particle interactions straightforward calculations yield
\begin{equation}
D_{0}(\rho)=\frac{\pi t}{L} \cos\left(\frac{\pi}{2}(1-\rho)\right) \text{cot}\left(\frac{\pi}{2L}\right)~,
\end{equation}
where $\rho=N/L$ is the filling, $N$ being the particle number.
In the following we calculate $D$ for the {\it interacting} Creutz ladder model, ${\cal H}_{0}+{\cal H}_{1}$, by using many-body diagrammatic perturbation theory first and an essentially exact numerical approach later.

The first-order contribution $E_{1}$ to the ground-state energy due to interactions is given by the Hartree and Fock diagrams~\cite{Pines_and_Nozieres,Giuliani_and_Vignale} in Fig.~\ref{HartreeFock}a) and~b): $E_1(\Phi) =L^{-1}\sum_{\sigma, k,k'} (W^{- - - -}_{k,k',0} - W^{- - - -}_{k,k',k'-k})$. 
After straightforward calculations, we find the first-order correction $D_{1} \equiv (L/2\pi) \left.(\partial^2 E_1 / \partial \Phi^2)\right|_{\Phi=0}$ to the Drude weight:
\begin{equation}\label{FirstOrderDrude}
D_{1} = \frac{\pi}{2L^2}\sum_{k,k'}  (g_{k} g_{k'} -f_{k} f_{k'} )
	 \left(v^{\sigma,\bar{\sigma}}_{k-k'} - v^{\sigma,\sigma}_{k-k'} + v^{\sigma,\sigma}_{0} - v^{\sigma,\bar{\sigma}}_{0} \right) .
\end{equation}
In writing Eq.~(\ref{FirstOrderDrude}) we have assumed that $v^{\sigma,\bar{\sigma}}_{q}$ and $v^{\sigma,\sigma}_{q}$ do not depend on $\sigma$ explicitly but just on the relative ``alignment'', as in the case of Eq.~(\ref{eq:interactions}). In general, the first-order contribution $D_1$ may yield an enhancement of the Drude weight,  i.e.~$D_{0} + D_{1} > D_{0}$ ~\cite{NegeleOrland}. The second-order contribution to the ground-state energy and associated Drude weight are instead negative definite. Direct inspection of Eq.~(\ref{FirstOrderDrude}) reveals that an interacting Creutz ladder model with on-site Hubbard-$U$ interactions only displays $D_{1}=0$. The same occurs for the case with $U=0$ and finite nearest-neighbor interactions, when same- and different-pseudospin fermions interact identically, i.e.~$D_{1}=0$ for $U=0$ and $v^{\sigma, \sigma}_{q} \equiv v^{\sigma,\bar{\sigma}}_{q}$. On the contrary, for $U=0$ and finite $V \neq \tilde{V}$, we find
\begin{eqnarray}
D_{1} &=&  \frac{\pi}{L^2}(V-\tilde{V})\frac{\csc^{2}(\pi/(2L))\sin^{2}(\pi \rho/2)}{1+2 \cos(\pi/L)}\nonumber\\
&\times&\cos(\pi/L)[\cos(\pi/L)-\cos(\pi\rho)]~.
\end{eqnarray}
It is easy to check that $D_{1}>0$ for $V> \tilde{V}$. 
\begin{figure}[t]
\centering
\begin{overpic}[width=\linewidth]{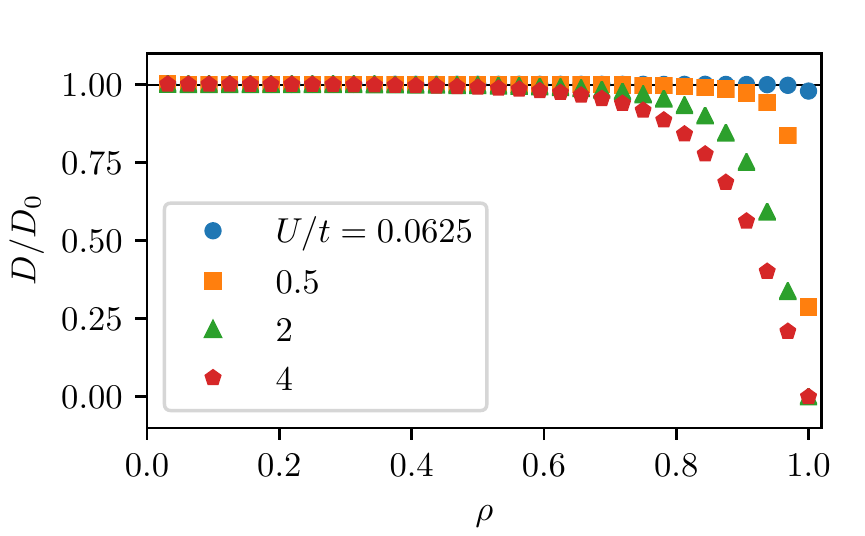}
\put(0,55){a)}
\end{overpic}
\begin{overpic}[width=\linewidth]{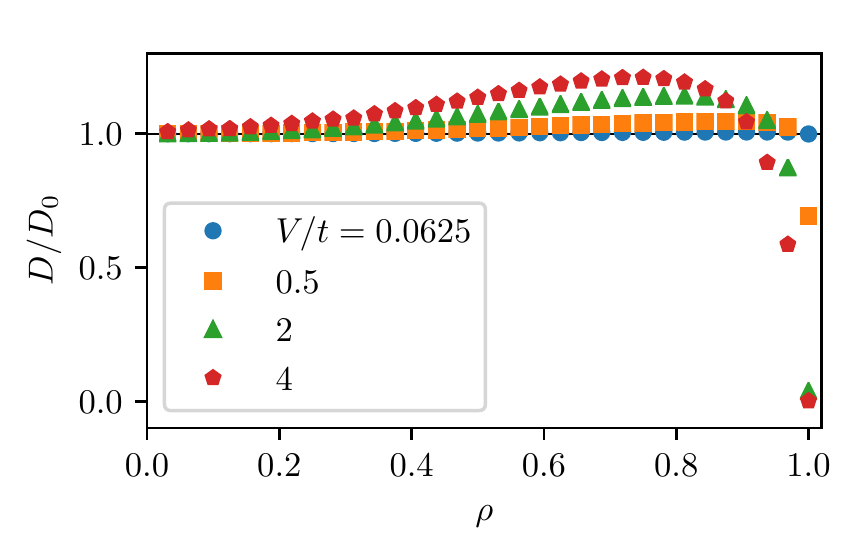}
\put(0,55){b)}
\end{overpic}
\caption{(Color online) Drude weight $D$ of a 1D interacting Creutz ladder model (with a Dirac-Weyl crossing for the choice $m=g=t$) measured in units of the non-interacting value $D_{0}$ and plotted as a function of filling $\rho = N/L$. 
Panel a) Drude weight {\it suppression} for finite values of $U/t$ and no nearest-neighbor interactions 
($V= \tilde{V}=0$). Panel b) Drude weight {\it enhancement} for $U/t=0$ and finite values of the nearest-neighbor interaction terms, keeping $V = 2 \tilde{V}$. Data in this plot refer to a chain with $L =32$ sites and $2\leq N\leq 32$. At half-filling, $D$ always tends to $0$ in the thermodynamic limit $L \to \infty$ (see text). The values of $D$ at all other fillings shown in these panels are instead fully converged at the aforementioned system sizes.\label{Fig3}}
\end{figure}
A brief summary of our second-order perturbation theory results is presented in Ref.~\onlinecite{SOM}. We emphasize that the first-order correction $D_{1}$ vanishes for non-relativistic 1D lattice models, for arbitrary interactions. 
For the Hubbard model, $f_k=g_k=$ constant, implying $D_{1}=0$.
A positive first-order correction $D_{1}$ to the Drude weight is a peculiarity of Dirac-Weyl fermions 
and seems directly linked to the ``spin-velocity locking'' occurring around the Fermi level. Indeed,
the form factors $N^\sigma_{k,s}$ are such that the spin of the eigenstates couples to their propagation direction.
Conversely, the Su-Schrieffer-Heeger model, which displays the same dispersion (i.e.~hosts a single massless Dirac fermion mode), does not have this feature and displays $D_1=0$.

\begin{figure}[t]
\includegraphics[width=\linewidth]{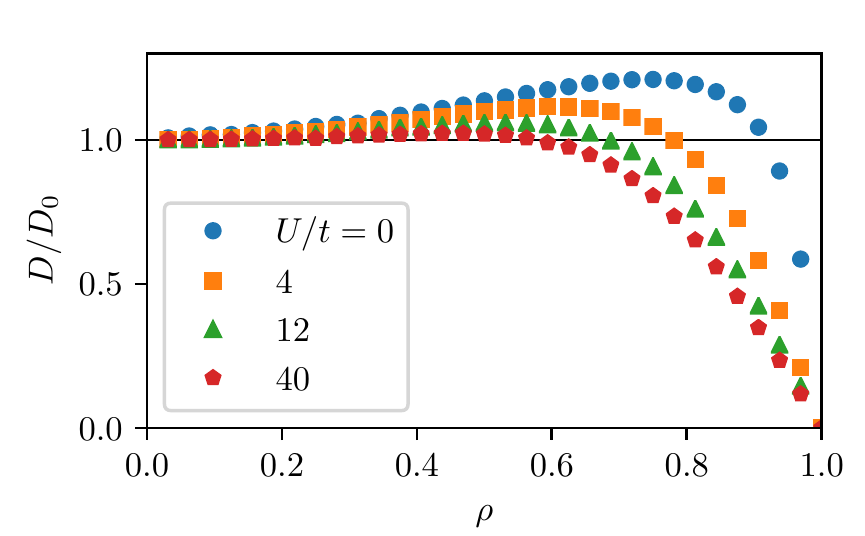}
\caption{(Color online) Stability of the Drude weight enhancement of an interacting Creutz ladder model ($m=g=t$) with respect to the value of the Hubbard on-site interaction $U/t$. The nearest-neighbor interaction parameters are fixed at $V/t=4$ and $\tilde{V}/t=2$. Data in this plot refer to a 1D ring with $L=64$ sites. The enhancement disappears only for $U \gg V, \tilde{V}$.\label{Fig4}}
\end{figure}

Guided by these analytical results, we now turn to a non-perturbative numerical analysis based on a tensor-network generalization of the DMRG approach~\cite{DMRG,cominotti_prl_2014}. Our main numerical results are summarized in Figs.~\ref{Fig3}-\ref{Fig4}.

We start by illustrating the dependence of $D$ on $\rho$ for a 1D Creutz ladder model with Hubbard-$U$ interactions only---see Fig.~\ref{Fig3}a). Results in this figure have been obtained by setting $m=g=t$, $V=\tilde{V}=0$, and finite $U$. Despite the Dirac-Weyl crossing in the single-particle band structure, this choice of parameters results in the usual scenario for non-relativistic Hubbard rings~\cite{Fye91,Scalapino92}, in which the Drude weight is suppressed by interactions, i.e.~$D/D_{0}<1$ for all fillings and coupling constants.

In stark contrast with the results in Fig.~\ref{Fig3}a), we cleary see that the Drude weight can display an entirely different qualitative behavior for finite nearest-neighbor interactions---see Fig.~\ref{Fig3}b). In this figure we show numerical results for the same system size ($L=32$), $m=g=t$, $U=0$, and $V=2\tilde{V}$. In this case, the Drude weight is {\it enhanced} (rather than suppressed) away from half filling, as expected on the basis of our analytical many-body perturbation theory analysis. The enhancement exceeds $20\%$ for $V/t=4$. At half filling, the usual Drude weight suppression takes place~\cite{Fye91,Scalapino92}, despite the different nature of inter-particle interactions. Of course, we have checked~\cite{SOM} that our numerical results in Figs.~\ref{Fig3} agree quantitatively with the many-body perturbation theory analysis for weak interaction strengths~\cite{SOM}.

{\it Stability of the Drude weight enhancement.---}Before concluding, we discuss the stability of the surprising result $D/D_{0}>1$ shown in Fig.~\ref{Fig3}b) with respect to finite values of $U/t$ and system size. 
Fig.~\ref{Fig4} shows the robustness of the Drude weight enhancement with respect to increasing values of $U/t$ and for a 1D ring with $L=64$ sites. We see that the enhancement persists at intermediate fillings for $U/t\neq 0$, disappearing only for $U \gg V, \tilde{V}$. Regarding the scaling of our results with respect to the system size $L$, we have checked that, away from half filling, the numerical data shown in Fig.~\ref{Fig4} are converged up to the second decimal digit.
Scaling towards the thermodynamic limit is slower at half filling. A finite-size scaling analysis of the ratio $D/D_{0}$ for this filling and rings with up to $L=128$ sites yields that even for $U/t=0$ and asymmetric nearest-neighbor interactions ($V=2\tilde{V}$) the Drude weight tends to zero in the thermodynamic limit~\cite{SOM}. This signals an interaction-induced metal-insulator transition, which is independent of the presence of the Dirac-Weyl crossing and range of inter-particle interactions.
In this work, we restricted our studies to the zero-temperature case.
For the case of finite temperatures, while the stability of the Drude weight in the thermodynamic limit $L\rightarrow\infty$ 
is still subject of debate~\cite{Sirker11}, signatures thereof should be anyway observable for finite systems in the low-frequency conductivity~\cite{Zotos2004}.

\noindent {\it Discussion.---}The realization of Dirac-Weyl crossings in 1D cold atom setups via the Creutz ladder model can be achieved by using available schemes~\cite{Mazza12,Juenemann16}. 
Given the crucial role of nearest-neighbour couplings, the natural systems to test our findings 
are dipolar atoms or molecules~\cite{Lahaye09,Carr09,Baranov12,Moses17}.
Recently, indeed, evidence for a long-range spin-exchange term~\cite{Yan13}, 
a deformation of the Fermi surface~\cite{Aikawa14}, and even of the realization of an extended Hubbard model~\cite{Baier16}, have been put forward.
Long-range interactions can also be achieved by dressing the natural ones via cavity resonances~\cite{Landig16} or lattice shaking~\cite{Eckardt16}. A precise scheme design goes however behind the scope of this work, and may be subject of future investigations. Incidentally, we note that previous theoretical studies of dipolar and Rydberg atoms on a ring~\cite{Olmos09,Zoellner11,Lim13} have focused on other aspects of their rich physics.
Our study therefore constitutes an additional motivation for the realization of such setups.
We also stress here that our findings are not limited to atoms loaded in ring traps.
The tunability of the Drude weight in the presence of Dirac-Weyl crossings is a much more general phenomenon, which can also be tested in transport experiments in cold atom wires, like the pioneering ones carried out at ETH~\cite{Brantut12}.
Another alternative could be to look at the long-time asymptotic
behaviour of current-current correlation functions, via tailored quenches~\cite{Karrasch17}.
\begin{acknowledgments}
{\it Acknowledgments.---}M.B. and M.R. acknowledge support from the Deutsche Forschungsgemeinschaft (DFG) through the grant OSCAR 277810020. 
J.J. thanks Studienstiftung des deutschen Volkes for financial support.
M.P. is supported by Fondazione Istituto Italiano di Tecnologia. 
The DMRG simulations were run by J.J. and M.R. on the Mogon cluster of the Johannes Gutenberg-Universit\"at 
(made available by the CSM and AHRP), with a code based on a flexible Abelian Symmetric Tensor Networks Library, developed in collaboration with the group of S. Montangero at the University of Ulm. 
\end{acknowledgments}

\newpage

\setcounter{page}{1}
\setcounter{figure}{1}

\appendix
\onecolumngrid
\hspace{5cm}

\section*{Supplemental Material for ``Tuning the Drude Weight of  Dirac-Weyl Fermions in One-Dimensional Ring Traps''}
\twocolumngrid
In this Supplemental Material file we present explicit calculations of the Drude weight of an interacting Creutz ladder model (with both on-site and nearest-neighbor interactions) based on second-order perturbation theory. The analytical predictions are compared with essentially exact numerical results, obtained by tensor networks simulations.
\section*{Second-order Perturbation Theory}
In the weak-coupling limit, we can use perturbation theory to approximate the ground-state energy of the interacting many-body system described by Eqs.~(1) and (4) in the main text. 

The Hugenholtz diagrams that contribute to the energy $E_{2} = E^{(1)}_{2} + E^{(2)}_{2}$ at second-order in the inter-particle interactions are shown in panels c) and d) of Fig.~2 in the main text.

These two diagrams can be calculated by using the Hugenholtz diagram rules (\cite{NegeleOrland} in main text). The Drude weight can then be obtained by differentiating the resulting energy expression twice with respect to the applied flux as in Eq.~(5) in the main text. 
We find

\begin{widetext}
\begin{equation}\label{eq:first-term}
E^{(1)}_2=\frac{1}{L^2}\sum_{k,k',q}\frac{1}{\varepsilon_-(\tilde{k}')-\varepsilon_+(\tilde{k}')}[ (k_-,k'_+|v|k_-,k'_-)- (k_-,k'_+|v|k'_-,k_-)][ (k'_-,q_-|v|k'_+,q_-)- (k'_-,q_-|v|q_-,k'_+)]
\end{equation}
and
\begin{equation}\label{eq:second-term}
E^{(2)}_2 =-\frac{1}{4L^2}\sum_{k,k',q}\frac{(k_-,k'_-|v|\bar{k}_+,\bar{k'}_+)- (k_-,k'_-|v|\bar{k'}_+,\bar{k}_+)}{\varepsilon_-(\tilde{k})-\varepsilon_+(\bar{k})+\varepsilon_-(\tilde{k}')-\varepsilon_+(\bar{k'}+2\varphi)}[ (\bar{k}_+,\bar{k'}_+|v|k_-,k'_-)- (\bar{k}_+,\bar{k'}_+|v|k'_-,k_-)]~,\\
\end{equation}
\end{widetext}
where $\bar{k}\equiv k-q$ and $\bar{k'} \equiv k+q$. For Hubbard-$U$ on-site interactions only, we have
\begin{equation}\label{eq:uterm}
(k_s,k'_{s'}|v|q_r,q'_{r'})= \frac{U}{2}\sum_{\sigma}N^\sigma_{\tilde{k},s}N^{\bar{\sigma}}_{\tilde{k}'s'}N^\sigma_{\tilde{q},r}N^{\bar{\sigma}}_{\tilde{q}'r'}~,
\end{equation} 
where the form factors $N^\sigma_{k,s}$ have been introduced in the main text.

\hspace{1cm}

For nearest-neighbor interactions, we find
\begin{eqnarray} \label{eq:vterm}
(k_s,k'_{s'}|v|&q_r,q'_{r'}) = \sum_{\sigma}[V \cos(k-q)N^\sigma_{\tilde{k},s}N^{\sigma}_{\tilde{k}'s'}N^\sigma_{\tilde{q},r}N^{\sigma}_{\tilde{q}'r'} \nonumber\\
&+\tilde{V} \cos(k-q)N^\sigma_{\tilde{k},s}N^{\bar{\sigma}}_{\tilde{k}'s'}N^\sigma_{\tilde{q},r}N^{\bar{\sigma}}_{\tilde{q}'r'}]~.
\end{eqnarray}
%

\begin{figure*}
\subfloat[]{
  \includegraphics[width=0.43\linewidth]{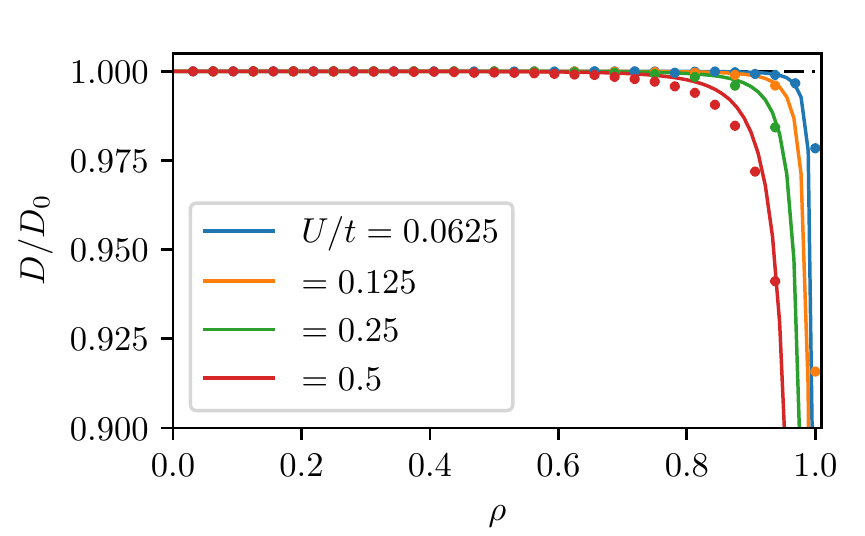}
}
\subfloat[]{
\includegraphics[width=0.43\linewidth]{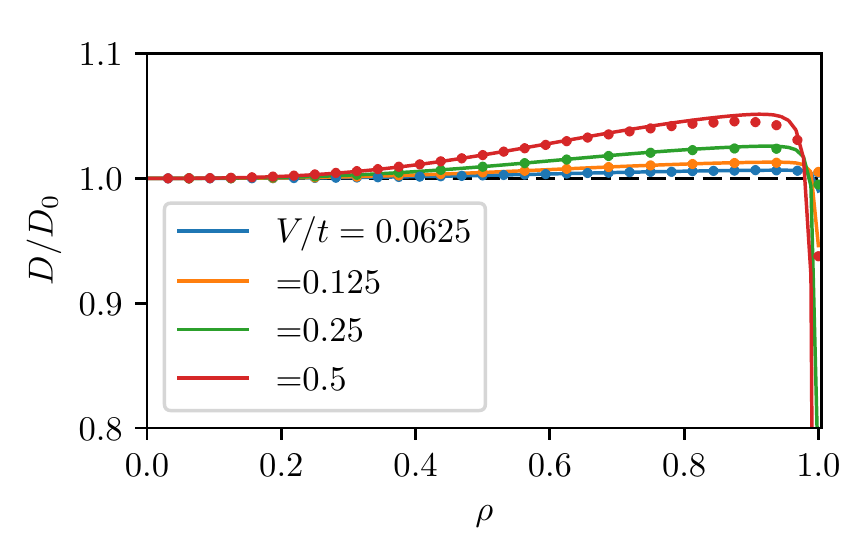}
}
\caption{\raggedright{}Analytical and DMRG results for weak-coupling interactions. Panel (a) On-site interactions $U$ (and $V=\tilde{V}=0$). Panel (b) nearest-neighbor interactions with $V=2\tilde{V}$ (and $U=0$). The solid lines are the analytical perturbation theory results in the thermodynamic limit, the data points show the DMRG results for a ring with $L=32$ lattice sites. \label{Fig1Supp}}
\end{figure*}

After some algebraic manipulations, we find that the second Hugenholtz contribution, Eq.~\eqref{eq:second-term}, always vanishes,
$E^{(2)}_{2}=0$. The second-order correction to the ground-state energy is finally
\begin{widetext}
\begin{eqnarray}\label{eq:important_result}
E_{2} = E^{(1)}_{2} = -\frac{1}{32L^2 t}\sum_{k'}\frac{1}{g_{\tilde{k'}}} && \left\{
U^2 \bigg[\sum_k\sin\left(\frac{k-k'}{2}\right)\bigg]^2 \right. \nonumber\\
&  &  + 2 \left.  \bigg[ \sum_k\sin\left(\frac{k-k'}{2}\right)
\left[(\tilde{V}+V)\cos(k-k')  + (\tilde{V}-V) (\cos\tilde{k} - \cos\tilde{k'} +1)  \right] \bigg]^2  \right\}~,
\end{eqnarray}
\end{widetext}
where  contributions of the different types of interactions are transparent.
Since $g_{\tilde{k}} =\text{sign}(\tilde{k}) \sin(\tilde{k}/2) \geq 0$, 
the second-order contribution to the energy is evidently always {\it negative} for any choice of the interaction strengths, as expected.

For purely on-site Hubbard-$U$ ($V=\tilde{V}=0$) or pseudospin-symmetric interactions, $V=\tilde{V}$ ($U=0$),
the only dependence of Eq.~\eqref{eq:important_result} on the flux $\varphi$ is via $g_{\tilde{k}}$.
Therefore, by elementary differentiation, it is evident that the second-order contribution to the Drude weight 
due to such interactions is also always {\it negative}.
This is confirmed by direct calculation, see Fig.~\ref{Fig1Supp}(a).
Recalling that the first-order correction vanishes identically for these two cases, $D_1 = 0$,
we conclude that the Drude weight of a 1D Creutz ladder model is suppressed by on-site Hubbard-$U$ interactions and symmetric nearest-neighbour interactions, $V=\tilde{V}$.

Analytical results of second-order perturbation theory and numerical DMRG results for the Drude weight 
for the case of Hubbard-$U$ on-site interactions only are compared in Fig.~\ref{Fig1Supp}(a). (Plots for $V=\tilde{V}$ look qualitatively similar.)
This is the standard result: repulsive on-site interactions suppress the Drude weight in comparison with the non-interacting value. 

Perturbation theory and numerical DMRG results for the Drude weight of the 1D Creutz ladder model with repulsive asymmetric nearest-neighbor interactions are shown in Fig.~\ref{Fig1Supp}(b). 

\section*{Finite-size scaling}
\begin{figure}[b]
\centering
\includegraphics[width=0.8\linewidth]{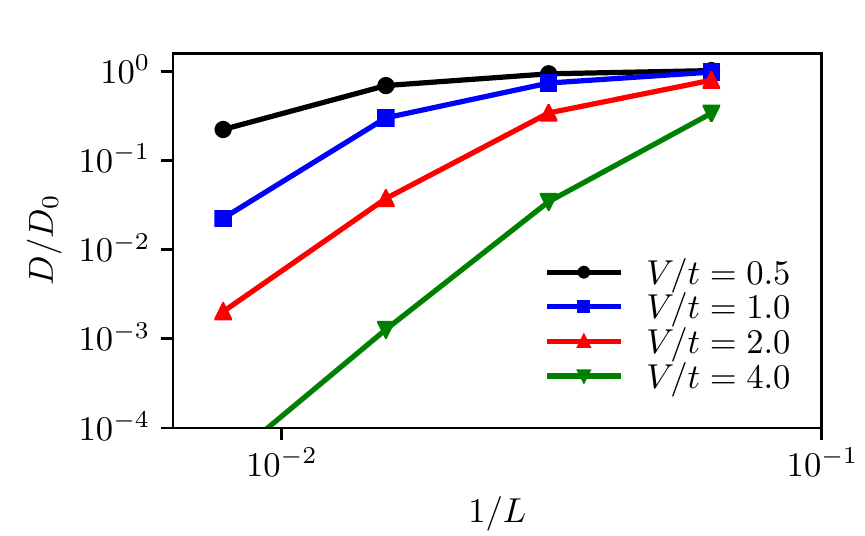}
\caption{\raggedright{}(Color online) Finite-size scaling analysis of the ratio $D/D_{0}$ at half filling for several values of $V$.
The largest system is a ring with $L=128$ sites. Data in this plot have been obtained by setting $m=g=t$, $\tilde{V}/t=2$, and $U/t=0$.}
\label{Fig5}
\end{figure}

A finite-size scaling analysis of the ratio $D/D_{0}$ for this filling and rings with up to $L=128$ sites is shown in Fig.~\ref{Fig5}. 
Even for $U/t=0$ and asymmetric nearest-neighbor interactions ($V=2\tilde{V}$) the Drude weight tends to zero in the thermodynamic limit.
This signals an interaction-induced metal-insulator transition, which is independent of the presence of the Dirac-Weyl crossing and range of inter-particle interactions.

\begin{thebibliography}{50}
\bibitem{paritynote}
Depending on the parity of the particle number, we evaluate $D$ at $\Phi=0$ for odd particle numbers 
or at $\Phi=\Phi_0/2$ for even particle numbers, to avoid spurious effects.
\end{thebibliography}
\end{document}